\documentclass[aps,prl,twocolumn,showpacs,amsmath,amssymb]{revtex4}
\usepackage{graphicx}
\usepackage{dcolumn}
\usepackage{bm}
\begin{document}

\title{Measuring Extreme Vacuum Pressure with Ultra-Intense Lasers}
\author{\'Angel Paredes$^1$, David N\'ovoa$^2$ and Daniele Tommasini$^1$}
\affiliation{$^1$ Departamento de F\'\i sica Aplicada,
Universidade de Vigo, As Lagoas s/n, Ourense, ES-32004 Spain;\\
$^2$ Centro de L\'aseres Pulsados, CLPU. Edificio M3 - Parque Cient\'{i}fico. Calle del Adaja, s/n, Villamayor, ES-37185 Spain.}

\begin{abstract}

We show that extreme vacuum pressures can be measured with current technology 
 by detecting the photons produced by the relativistic Thomson
scattering of ultra-intense laser light by the electrons of the medium. 
We compute the amount of radiation scattered at different frequencies and 
angles and design strategies for the efficient measurement of pressure.
In particular, we show that a single day experiment at a high repetition rate Petawatt laser facility such as VEGA, that will be operating in 2014 in Salamanca, will be sensitive, in principle, to pressures $p$ as low as $10^{-16}$Pa, and will be able to provide highly reliable measurements for  $p\gtrsim10^{-14}$Pa.

\end{abstract}

\pacs{42.62.-b, 07.30.Dz, 41.60.-m, 52.38.-r}

\maketitle


{\em Introduction.-} 
Pressures corresponding to Extreme-High Vacuum (XHV), $p<10^{-10}$Pa \cite{redhead98}, are measured by
ionization methods: the atoms in the sample are ionized and the produced charged particles
are collected by applying an electric field. This procedure is fully reliable for pressures as low as $10^{-11}$Pa \cite{redhead}. Although there are techniques able to push  
this limit down\cite{Chen87}, its use would be questionable since in this regime the electron stimulated desorption, the so-called X-ray limit, or the out-gassing from the hot cathode cannot be neglected \cite{redhead,calcatelli}. It is therefore of crucial importance the introduction of a new, alternative method aimed at providing an independent measurement of XHV pressures below $10^{-11}$Pa without significantly altering the pressure itself and free of the aforementioned limitations. 

In this Letter, we propose the idea of using photons to gauge the extreme 
vacuum properties. For this purpose, the advent of ultra-high intensity
lasers \cite{mourou06} has provided a new class of light sources which are powerful enough to produce a measurable signal even in conditions of XHV. In the XHV, the remnant pressure is essentially produced by the hydrogen
released by the walls. When interacting with high-intensity laser light, the electrons
can be considered as free, therefore the main source of dispersed light is relativistic
nonlinear
Thomson scattering. 
This process was studied in detail in \cite{Eberly,sarachik,castillo,esarey,salamin} and experimentally observed
in \cite{experiment}, and may be used to measure the peak intensity of a laser pulse \cite{peak}.

We will compute the number of scattered photons as a function of the electron density of the medium and the parameters of the ultra-intense laser pulse (wavelength, peak power, waist radius, pulse duration
and repetition rate). 
In the XHV regime, collective effects of the electrons as those discussed in
\cite{castillo,esarey} can be neglected. The
number of scattered photons is proportional to the number of scattering centers, which is proportional to the pressure. Harmonics are generated in the scattering process. It will be shown that most of the scattered photons correspond to the incident wavelength ($n=1$). Nevertheless, detection of photons with $n=2,3,\dots$ may be
possible and useful.
Remarkably, we find that
it should be possible to provide a highly reliable measurement of pressures as low as $p \approx 10^{-14}$ Pa
in realistic
conditions at facilities that will be available in the near future.
As a consequence, ultra-intense lasers may be able to push the physical limits for measuring a basic magnitude like pressure. A lower cost practical application of this result can be the use of more common intermediate-intensity lasers as an alternative instrument of measuring high vacuum pressure in a non-extreme regime. 

On the other hand, ultra-intense lasers need high vacuum to operate and it is important to determine their operation conditions. Moreover, extreme vacuum is a necessary requirement of many of the experiments based in ultra-intense lasers that have been proposed in the last few years aimed at demonstrating the quantum vacuum polarization\cite{PPSVsearch,NaturePhotonics2010,diffraction} and at searching for new particles\cite{new_physics}. It is fascinating in these cases that the laser itself may provide an efficient tool to monitor
the pressure in the chamber, substituting or complementing other conventional methods.

{\em Relativistic Thomson scattering.-} 
Our computations are based on some of the results of \cite{sarachik}, which we briefly
review for completeness and to fix notation. We introduce
the dimensionless parameter $q$, related to
the intensity $I$ and wavelength $\lambda_0$ of the beam as:
\begin{equation}
q^2 = \frac{2I\,r_0 \lambda_0^2}{\pi \,m_e c^3},
\label{q2}
\end{equation}
where $r_0\approx 2.82 \times 10^{-15}$m is the classical electron radius.
Let us also define ${\cal M}=1+\frac12 q^2 \sin^2(\theta/2)$.
Relativistic effects play a role for $q\gtrsim 1$, corresponding
to $I \gtrsim 2\times 10^{18}$W/cm$^2$ (for $\lambda_0=800$nm).
When a linearly polarized 
plane wave impinges
on a free electron, 
the power scattered per unit solid angle is:
$\frac{dP^{(n)}}{d\Omega}= \frac{e^2 c}{8\epsilon_0 \lambda_0^2} 
f^{(n)}$,
where $n$ is the harmonic number and $f^{(n)}$ is dimensionless:
\begin{eqnarray}
f^{(n)}=\frac{q^2n^2}{{\cal M}^4}
\Bigg[\left( 1-\frac{(1+\frac12 q^2) \cos^2\alpha}{{\cal M}^2} \right)(F_1^n)^2+\nonumber\\
-\frac{q\,\cos\alpha (\cos\theta - \frac12 q^2 \sin^2(\theta/2))}{2{\cal M}^2}
F_1^n F_2^n+\frac{q^2 \sin^2 \theta}{16 {\cal M}^2 }(F_2^n)^2\Bigg],
\label{fn}
\end{eqnarray}
with $\cos\alpha= \sin\theta \,\cos \varphi$, where
$\theta \in [0,\pi]$, $\varphi \in [0,2\pi)$ are usual spherical coordinates. 
Forward scattering corresponds to $\theta=0$ and
$\varphi=0,\pi$ point along the polarization axis.
The $F_s^n$ can be written in terms of Bessel functions as:
$F_s^n = \sum_{l=-\infty}^{+\infty}J_l\left(\frac{n\,q^2\sin^2(\theta/2)}{4{\cal M}}\right)
\times \nonumber\\
\left[J_{2l+n+s}\left(\frac{q\,n\,\cos\alpha}{{\cal M}}\right)+
J_{2l+n-s}\left(\frac{q\,n\,\cos\alpha}{{\cal M}}
\right)
\right]$.
These results hold in the laboratory frame, in which the scattered wavelength is shifted as $\lambda^{(n)}=
{\cal M}\lambda_0 / n$.

In this paper, we will only consider linearly polarized laser pulses. The main difference in the case of circular polarization would be the lack of the dependence on the azimuthal angle. 

{\em Modeling a realistic situation.-} 
Our main goal is to compute the average 
number of photons scattered when a laser pulse traverses a vacuum chamber.
It is crucial to take into account the distribution in space of the incident radiation.
We model the pulse as having a Gaussian profile (waist radius $w_0$). 
The intensity then reads:
$
I=I_0  \left(\frac{w_0}{w(z)}\right)^2 e^{-\frac{2r^2}{w(z)^2}}
$.
The beam radius evolves as $w(z)=w_0\sqrt{1+z^2/z_R^2}$ where
the Rayleigh range is $z_R = \pi\,w_0^2/\lambda$. 
Our estimate for the scattered radiation will
be obtained by performing the appropriate integral after inserting 
the intensity profile in (\ref{fn}). In particular, the number of photons of the $n$'th harmonic
produced by a single pulse
are $N_\gamma^{(n)}=n_e \tau \int  \frac {dP^{(n)}}{d\Omega} 
\frac{{\cal M} \,\lambda}{h\,c\,n}  d\Omega d^3 \vec x$,
where $n_e$ is the number of electrons per unit volume, which we will assume to be uniform
and $\tau$ the pulse duration. It is useful to rewrite the integral in terms of the
dimensionless quantities $\rho \equiv r/w_0$,
$\xi \equiv z/z_R$, such that:
\begin{equation}
q^2 = q_0^2 \frac{1}{1+\xi^2}\exp\left(-\frac{2\rho^2}{1+\xi^2}\right),
\label{qpos}
\end{equation}
where $q_0$ is related to the intensity at the beam focus.
 After some simple manipulations one can write:
\begin{equation}
N_\gamma^{(n)}= {\cal K} \int
\int \rho\,\Gamma^{(n)}(q)   d\rho d\xi,
\label{Ngamma}
\end{equation}
where we have introduced a function of $q$:
\begin{equation}
\Gamma^{(n)}(q)=\int_0^{2\pi}\int_0^\pi \frac{1}{n}f^{(n)} {\cal M} \sin\theta d\theta d\varphi
\label{Gammaint}
\end{equation}
and the parameter:
\begin{equation}
{\cal K}=\frac12 n_e (c\,\tau) \frac{\pi^2 w_0^4}{\lambda_0^2} \alpha,
\end{equation}
 where
$\alpha \approx \frac{1}{137}$ is the fine structure constant. 
 Notice that the integral in Eq. (\ref{Ngamma}) only
depends on $n$ and on $q_0$ whereas the rest of quantities describing the physical
situation are factored out
in ${\cal K}$.

\begin{figure}[htb]
{\centering \resizebox*{\columnwidth}{0.5\columnwidth}{\includegraphics{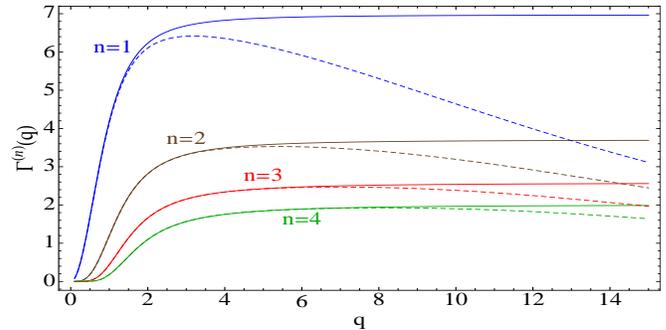}}} 
\caption{(Color online) In solid lines, the function $\Gamma^{(n)}(q)$ found by numerical integration.
Dashed lines are found by cutting the $\theta$-integration in Eq. (\ref{Gammaint})
with $\theta_{cut}=0.1$.
}
\label{figGamma}
\end{figure}

We have computed the number, frequency and spatial distribution of the 
photons
that may be produced as a function of the incoming laser pulse parameters,
by numerical integration of the expressions in
(\ref{Ngamma}), (\ref{Gammaint}).
Fig.
 \ref{figGamma} shows a plot of the function $\Gamma_n(q)$  for $n=1,\dots,4$. 
Since it is impossible to have a detector covering the full
 solid angle, we also show the results when  the integral is
 performed over a 
 reduced range of the polar angle $\theta \in (\theta_{cut},\pi-\theta_{cut})$. 
In any realistic situation one should have $\theta_{cut} \gg \theta_d$, being
 $\theta_d=\lambda_0/(\pi\,w_0)$ the beam divergence. 

\begin{figure}[htb]
{\centering \resizebox*{\columnwidth}{0.5\columnwidth}{\includegraphics{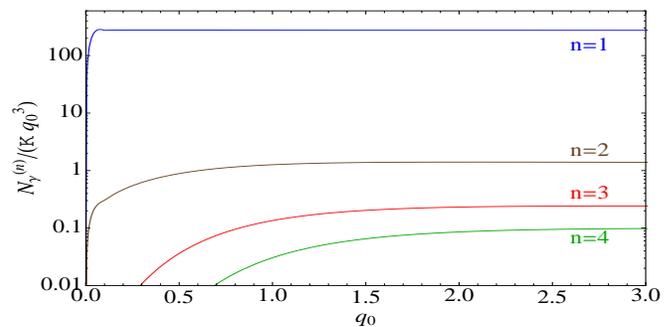}}} 
\caption{(Color online) Semi-logarithmic plot of $N_\gamma^{(n)}/({\cal K}\,q_0^3)$ for $n=1,\dots,4$.
}
\label{figN}
\end{figure}

Using these results for $\Gamma^{(n)}(q)$ and the Gaussian intensity distribution
(\ref{qpos}) one can readily compute the integral in (\ref{Ngamma}). 
We have confined the integration to the region where an electron of a hydrogen
atom can be considered as free $q>q_{cut}= 0.01$, see discussions below for more details. 
The results are plotted in
Fig. \ref{figN}. 
We find the following asymptotic behavior (valid for $q_0$ large enough depending on $n$):
\begin{equation}
N_\gamma^{(n)} \approx c_n {\cal K}\,q_0^3 ,
\label{Ngresult}
\end{equation}
with the values $c_1\approx 275$, $c_2\approx 1.3$, $c_3\approx 0.22$, $c_4\approx 0.088$. 
If for Fig. \ref{figN}
 one performs the same cut as before in the integration region
 $0.1 < \theta < \pi - 0.1$, the correction to the result is tiny, below 1\%.
 This happens because most of the photons are {\it not} generated at the maximum intensity region
 --- the beam focus ---, but at the larger volume where the Gaussian profile presents moderate
 values of $q$, irrespective of how large $q_0$ might be. In particular, for $n=1$ most of the photons are generated in a region with $q<1$ and therefore
 one can find an approximation to the $n=1$ result using the simpler expressions for non-relativistic
 Thomson scattering. One gets $c_1 \approx 8\pi/(9 q_{cut})$.
 
 In order to understand qualitatively the results for $n>1$, we may approximate
 the plateaus of $\Gamma^{(n)}(q)$ displayed in figure \ref{figGamma} by Heaviside
 step functions $\Gamma^{(n)}(q)= b_n \Theta(q-q_{step,n})$. Then, defining the limits of
 the $q>q_{step,n}$ region as 
  $\rho_{lim}=\sqrt{\frac{1+\xi^2}{2}\log\left(\frac{q_0^2}{q_{step,n}^2(1+\xi^2)}\right)}$
 and $\xi_{lim}= \sqrt{\frac{q_0^2}{q_{step,n}^2}-1}$, one can
 estimate the integral in
  Eq. (\ref{Ngamma}) as $\int_{-\xi_{lim}}^{\xi_{lim}}\int_0^{\rho_{lim}} \rho\,b_n d\rho d\xi
  \approx \frac{b_n q_0^3}{9q_{step,n}^3}$ where we have only kept the leading
  term in $q_0/q_{step,n}$. This simplified analysis 
  fits qualitatively the 
    numerical results and explains
  the cubic dependence in $q_0$: it is a consequence of the fact that, roughly, both 
 $\rho_{lim}$, $\xi_{lim}$
   grow linearly with $q_0$.
   
\begin{figure}[htb]
  {\centering \resizebox*{0.48\columnwidth}{0.5\columnwidth}{\includegraphics{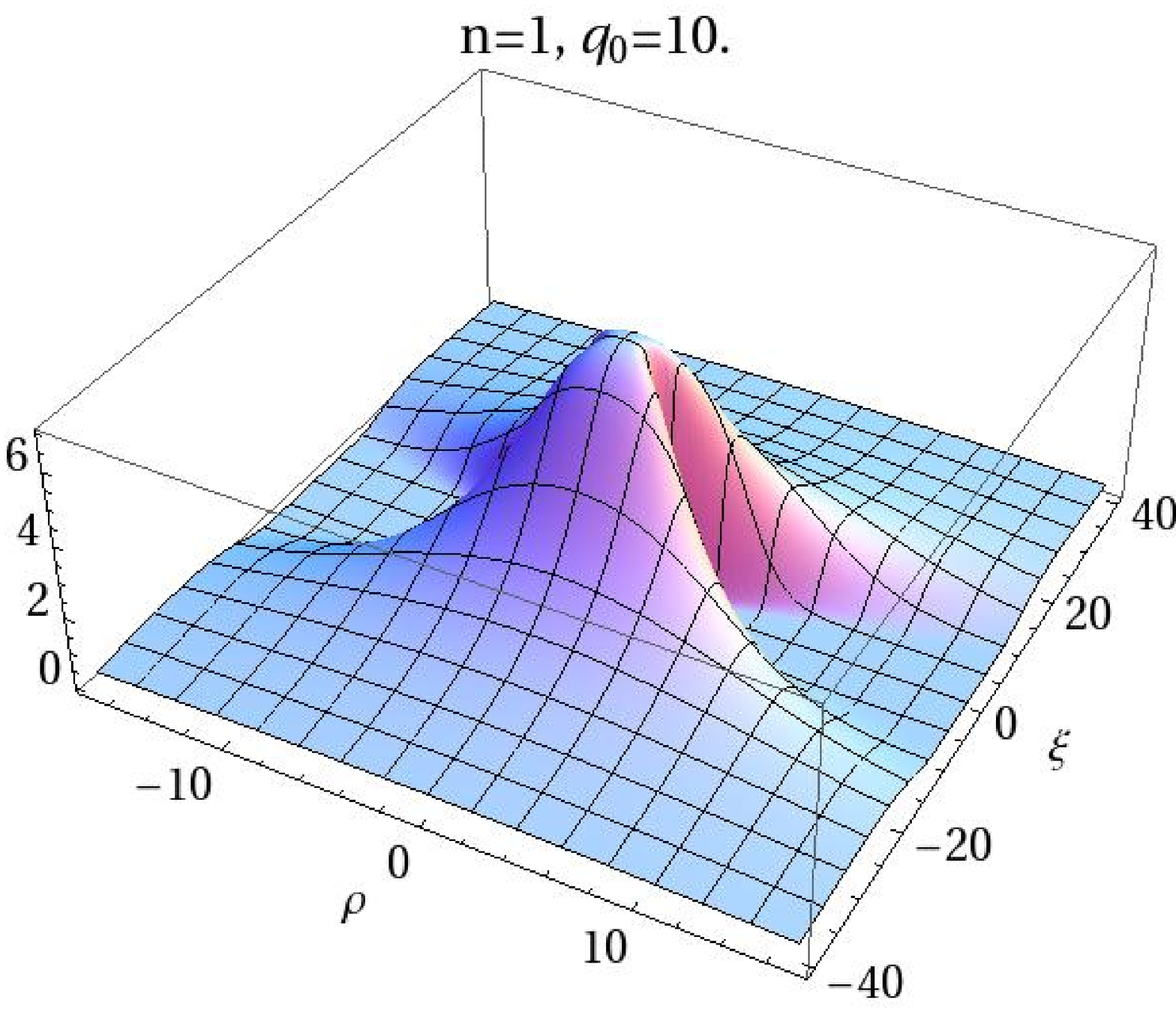}}
  \resizebox*{0.48\columnwidth}{0.5\columnwidth}{\includegraphics{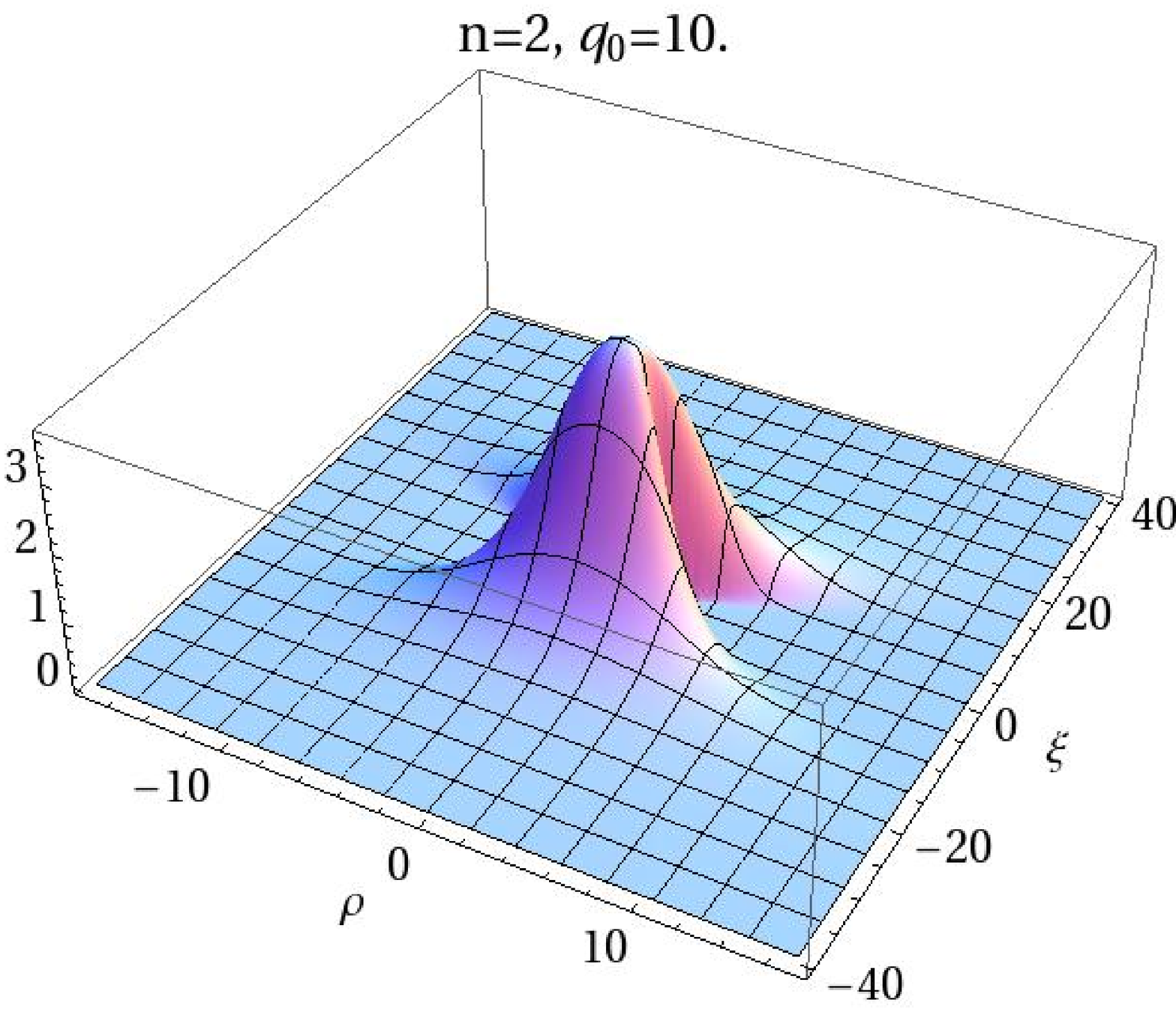}}} 
  \caption{(Color online) $\Gamma^{(n)}(q(\rho,\xi))$ in two cases. For $n>1$, photons are produced in
  the region where $q\gtrsim 1$. The region where $n=1$ photons are scattered is larger
  since they are the only outcome of the non-relativistic regime $q\ll 1$.
  }
  \label{figrhoxi}
  \end{figure}

  In Fig. \ref{figrhoxi}, we present  plots of the value of
  $\Gamma^{(n)}(q(\rho,\xi))$, which show the region in space in which the  photons
  are scattered.
  The information displayed in Fig. \ref{figrhoxi} indicates which the
   precise region of the 
  chamber where the electron density measurement is taking place.
  
\begin{figure}[htb]
  {\centering \resizebox*{0.48\columnwidth}{0.5\columnwidth}{\includegraphics{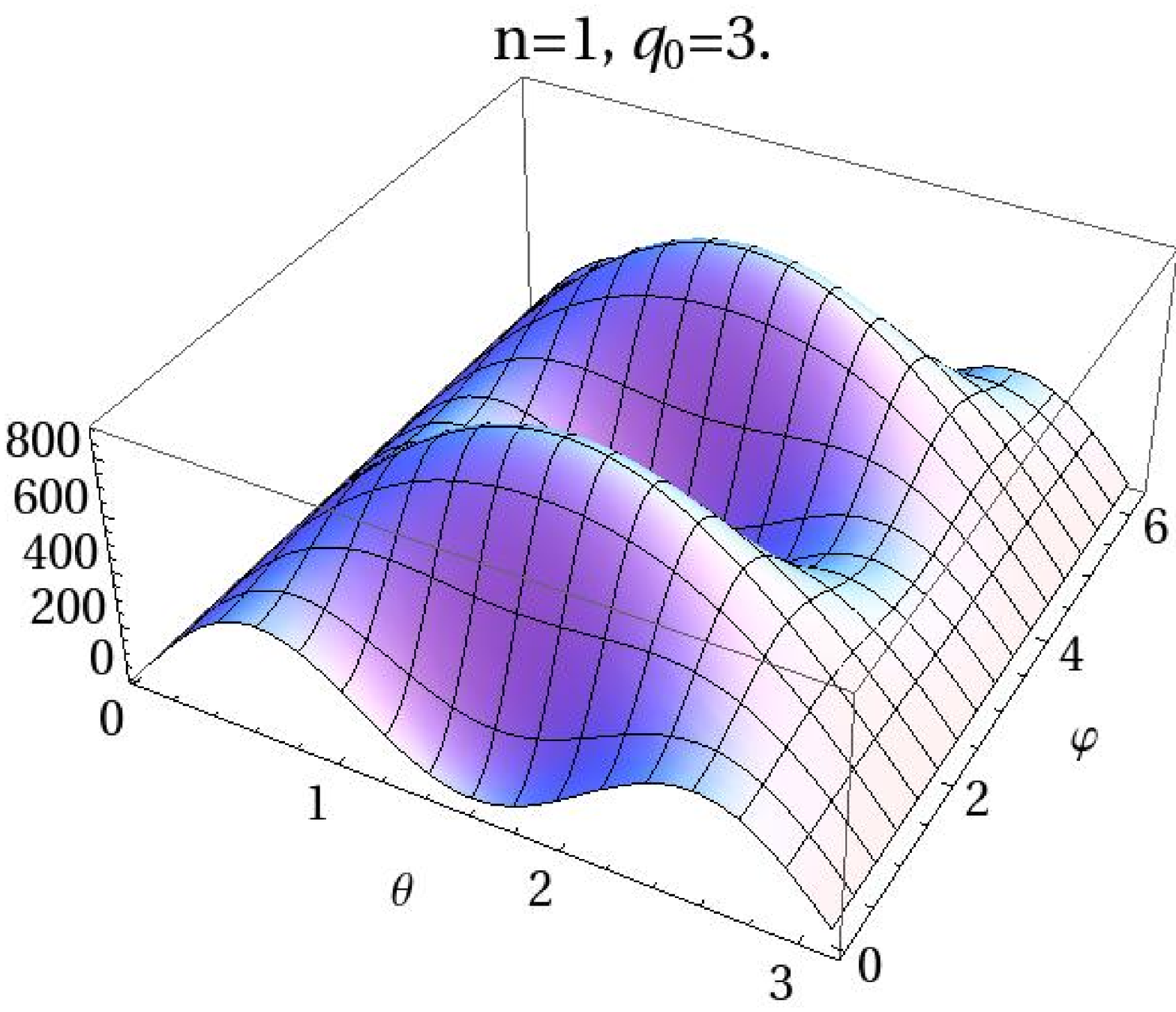}}
  \resizebox*{0.48\columnwidth}{0.5\columnwidth}{\includegraphics{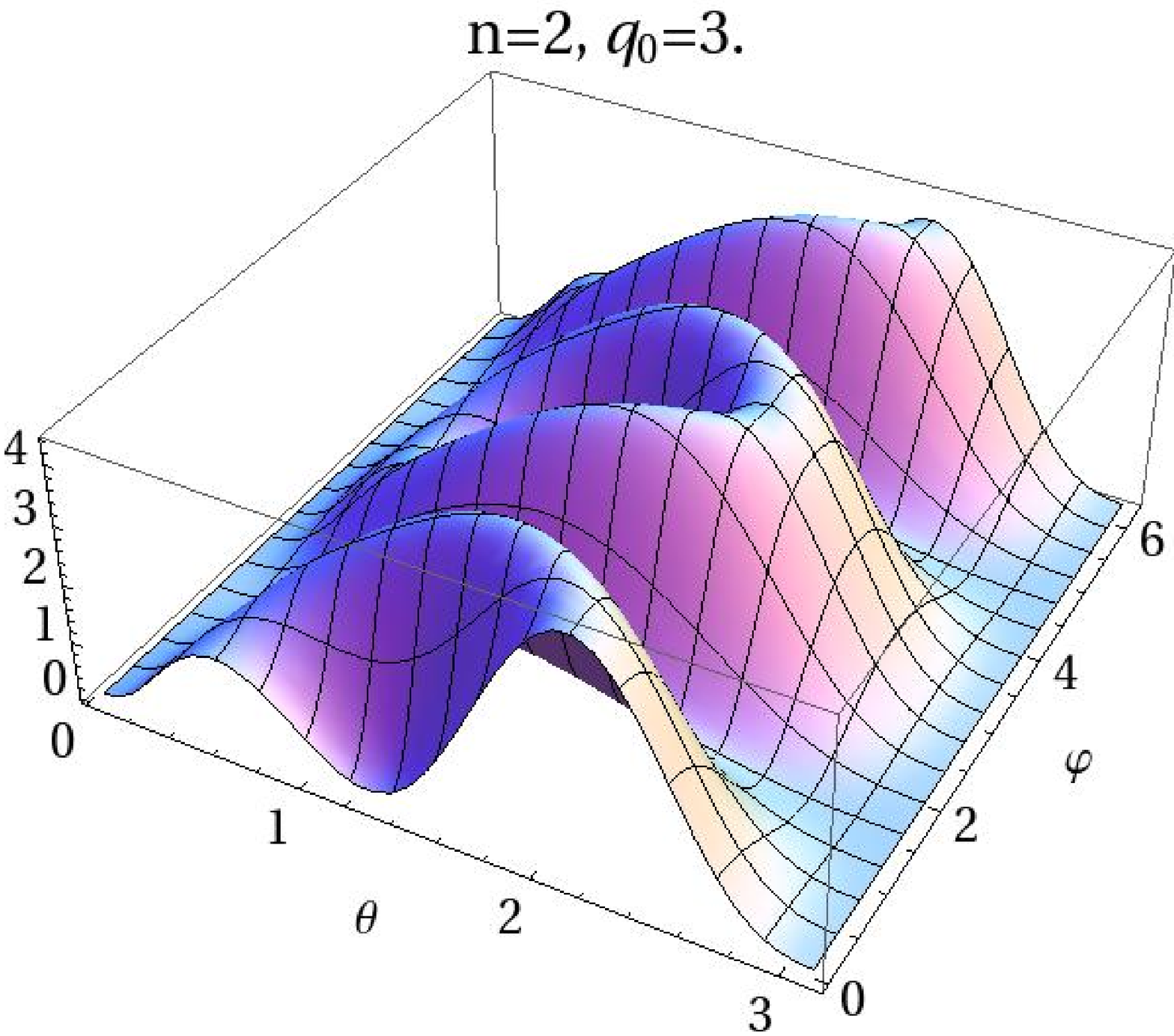}}} 
  \caption{(Color online) Plots of ${1}/n \int \int \rho 
  f^{(n)} {\cal M} \sin \theta d\rho \,d\xi$ as a function of 
  $\theta$ and $\varphi$, see Eqs. (\ref{Ngamma}), (\ref{Gammaint}).}
  \label{figangular}
  \end{figure}

In Fig. \ref{figangular} we plot two examples of
the angular distribution of the emitted photons, obtained by performing the integral in (\ref{Ngamma}) on the
  $\rho-\xi$ space and leaving it as a function of $\theta$ and $\varphi$. 
For a given harmonic,
the angular dependence does not change too much when modifying $q_0$. The reason for this is
 that --- as noted above ---
even when $q_0$ is large, a copious amount of radiation comes from the region of smaller $q$. 
This same argument explains why the distributions are not forward peaked, as one
may naively expect. In fact, the plot for $n=1$ can be hardly distinguished from
the angular distribution corresponding to linear Thomson scattering $\sin\theta(1-\sin^2\theta
\cos^2\varphi)$.
These considerations may be
useful when looking for an optimised configuration of photon detectors in an actual experiment.

\begin{figure}[htb]
  {\centering \resizebox*{0.95\columnwidth}{0.5\columnwidth}{\includegraphics{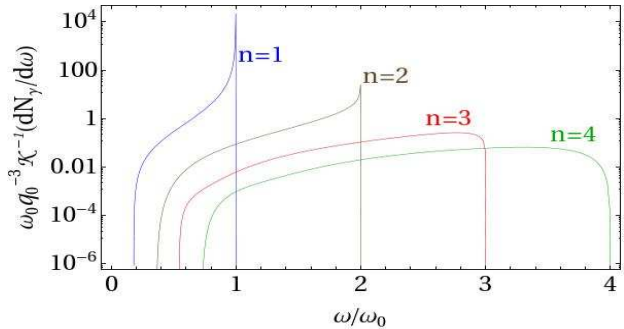}}} 
  \caption{(Color online) Differential frequency distribution of the scattered radiation
  for $n=1,\dots,4$, in logarithmic scale. }
  \label{figspectrum}
  \end{figure}

We now turn to the frequency distribution of the scattered photons.
Harmonics are not emitted in multiples of the original frequency but there is a shift
related to electron recoil $\lambda^{(n)}={\cal M}\lambda_0 / n$, where ${\cal M}$ depends
on $q$ and $\theta$. Formally, we can find the spectral distribution by writing
$\frac{d N_\gamma^{(n)}}{d\lambda^{(n)}}= {\cal K} \int d\rho \int d\xi 
\int d\theta \int d\varphi \left(\rho\,\frac{1}{n}f^{(n)} {\cal M} \sin\theta\,
\delta\left(\lambda^{(n)}- \frac{{\cal M}\,\lambda}{n}\right)
\right)$.
In Fig. \ref{figspectrum}, we plot the result of the numeric integration for
$q_0=3$. The result does not appreciably change for larger values of $q_0$. 
In the final part of this Letter, we will comment on limitations of this computation.

{\em Quantitative estimates.-} 
Since the number of scattered photons is directly proportional to the number of scattering
centers,
at extreme low pressure, the scattered radiation will be extremely weak. 
In order to get an idea of whether a given pressure can be measured with
this method, we must know the number of photons that can be detected in a reasonable amount
of time. 
If we assume that only atomic hydrogen is left in the chamber, the relation
between the electron density and the pressure is $n_e=p/(k_B T)$.
Using the result (\ref{Ngresult}), we can estimate the total number of
photons detected in a period of time $\Delta t$ in terms of the laser repetition rate $r_r$,
the total energy of each pulse $E_{pulse} = \tau\, I_0 \pi \,w_0^2/2$ and the detector efficiency
$f$ --- which include geometric and quantum factors. 
\begin{equation}
N_{\gamma,det}^{(n)} \approx \frac{4c_n}{\pi} (\Delta t\,r_r)f\frac{p}{k_B T}
 \alpha\,\frac{w_0 \lambda_0 r_0^{3/2}}{(c\,\tau)^{\frac12}}
\left(\frac{E_{pulse}}{m_e c^2}\right)^{\frac32},
\label{Ngresult2}
\end{equation}
It might seem weird that for fixed energy the signal grows with the waist radius.
What happens is that having smaller maximum intensity is compensated by a larger
interaction region. Nevertheless, when $w_0$ it too large, $q_0$ becomes small
and (\ref{Ngresult}) and
(\ref{Ngresult2}) lose their validity (see Fig. \ref{figN}).

Eq. (\ref{Ngresult2}) is the main result of this paper. To be concrete, we can now evaluate the limiting pressure that can be measured with this method at a given ultra-intense laser facility in a reasonable time span. As an example, we will consider the Petawatt laser VEGA that will be available in 2014 at the CLPU of Salamanca \cite{VEGA}, having repetition rate as large as $r_r=1s^{-1}$, 
with pulses of $\lambda_0=800$nm, $\tau=30$fs and $E_{pulse}= 30$J. 
Taking e.g. $T=300$K, $w_0 = 20 \mu$m and a day run, $\Delta t = 1$ day, and assuming an efficiency $f=0.5$  for $n=1$ 
which is a realistic value for commercially available single photon detectors at $\lambda_1=800$nm, we can compute the limiting pressure that can be measured within 3 standard deviations by taking $N_{\gamma,det}^{(1)} =10$ in Eq. \eqref{Ngresult2}. We obtain $p_{\rm limit}\simeq 10^{-16}$Pa. Of course, this sensitivity should be corrected by a geometric efficiency factor,  depending on the effective area of the detector that is chosen.
Geometric efficiency corrections will be more important for the $n=1$ photons since the region where
they are scattered is large, Fig. \ref{figrhoxi}.
 Note that the angular cut that we have imposed in our computation ensures that the detected photons will not be confused with those of the beam that do not undergo Thomson scattering, that would give no observable signal in the integration area for one day run (other sources of noise will be considered below).
Another way of avoiding such kind of background would be the measurement of the $n=2$ harmonic. From Eq.  (\ref{Ngresult2}), we obtain that the limiting pressure that can be measured by detecting $N_{\gamma,det}^{(2)} =10$ photons after one day run would be $p_{\rm limit}\simeq 10^{-14}$Pa, if all the other parameters are taken as above except the efficiency, that can be as large as $f\simeq0.6$ for $\lambda_2\simeq400nm$ in state-of-art single photon detectors. For all these reasons, we conclude that the detection of the $n=2$ harmonic will provide an independent measurement of the pressure above the $10^{-14}$Pa range for a 1 day run at VEGA, complementary to the more sensitive measurement due to the $n=1$ wave. Taken together, these measurements could be used for a kind of self-calibration of the whole procedure. The result for the measurement of pressure as low as $10^{-14}$ Pa would then be highly reliable, 
provided that the detection is accurate enough and the noise level can be kept below the signal, which are feasible tasks with present technology as we discuss below. 

{\em Background analysis.-} A source of noise are thermal photons, whose expected number per shot is
\begin{equation}
N_{\gamma,{\rm det}}^{{\rm th}}\simeq 2\pi\, c  S_{\rm det} \Delta t \int d\lambda \frac{f(\lambda) {\lambda^{-4}}}{1+\exp\left(\frac{hc}{k_B T \lambda}\right)},
\label{thermal_back}
\end{equation}
where $f(\lambda)$ is the sensitivity of the detector as a function of the wavelength $\lambda$, 
$S_{\rm det}$ is the detecting area and $T$ is the temperature of the vacuum tube. At ordinary temperatures $T\simeq 300K$, this background can be made completely negligible in all the configurations that are of interest for the present work by using a wavelength filter on the detector, cutting off all the wavelengths larger than $\lambda_0+\Delta\lambda_0$, where $\Delta \lambda_0\simeq\lambda_0^2/(c\tau)$ is the uncertainty in the pulse wavelength. 

A potentially higher source of noise is due to the dark counts of the detector, that can be kept below $10 s^{-1}$ in avalanche photodiodes featuring high efficiencies 
at the wavelengths discussed in this Letter. If we require that  during each repetition the detection window is opened for a very short time, which can be as short as two nanoseconds with present technology, we can ensure that after the $\sim10^5$ repetitions in the 1 day experiment at VEGA the total dark count would be unobservable. Such gating of the detector would also provide an efficient protection mechanism against the backscattered photons from the walls of the vacuum tube. In fact, assuming space dimensions of the tube of the order of few tens of centimeters or larger, the backscattered photons would reach the detector out of the detection window. A promising alternative could also be the use of superconducting single photon detectors\cite{detectors1,detectors2}, that are able to reduce the dark counts below $10^{-2}s^{-1}$ in both the visible and infrared ranges.

{\em Validity of approximations.-}
We discuss now several approximations and assumptions that have been made in deriving
our results.
First of all, radiation reaction and quantum effects have been
neglected in Eq. (\ref{fn}), which is a good approximation since $q^2 \ll \lambda_0/r_0$ and
$n\,h\,c/\lambda_0 \ll m_e c^2$ in relevant situations. 
Moreover, (\ref{fn}) is valid 
for a plane wave. This means that the Gaussian beam radius should be larger than the transverse
displacement of the electron which is typically of order $q\,\lambda$. Namely, the formalism
is valid for $w_0 \gg q_0 \lambda_0$ and cannot be used for diffraction limited beams. 
We have considered the radiating electrons as free. This is valid as long as the atomic potentials
can be neglected in the presence of the laser beam, namely in the barrier suppression regime
\cite{barriersup}, which
for hydrogen corresponds to $I>I_{BS}=1.4 \times 10^{14}$W/cm$^2$
\cite{barriersupH}. Since this limiting value corresponds to $q \ll 1$
(it is $q \approx 0.01$ for $\lambda_0=800$nm)
 --- and $q\approx 1$ is related to the onset of relativistic effects---, we conclude that 
the binding energy of the electrons does not play a role in the harmonic generation we have discussed.
For the same reason, harmonic generation coming from electron-proton recombination is suppressed
(and would be further suppressed if polarization is non-linear) leaving
relativistic Thomson scattering as the dominant process. 
It should also be mentioned that
 we have used expressions for electrons initially at rest since at room temperature they
are far from being relativistic.

Finally, we have made a rough modeling considering square pulses and 
not taking into account the effects of the time envelope of the pulses. 
For short pulses --- with a small or moderate number of light cycles --- this
approximation is not accurate, as it has been discussed both in classical
\cite{krafft,Gao} and quantum \cite{mackenroth} frameworks.
The main correction that will appear is a spectral broadening
--- called 
{\it ponderomotive broadening} in \cite{krafft}. Because of this phenomenon and of the fact that the initial
 beam is not monochromatic, the plots of Fig. \ref{figspectrum} might markedly underestimate
 the spectral width of the produced harmonics for short pulses.
Corrections may also multiply Eq. (\ref{Ngresult2})
by a factor of order 1, but we do not expect
 them to change the orders of magnitude. However, it could be pertinent to make a full study when
 dealing with a particular situation. 

{\em Conclusions.-} We have computed the amount
and spectral distribution of the photons that are produced when a
gaussian laser pulse crosses a vacuum tube. With present detector and ultra-intense laser technologies, this implies the possibility of measuring pressures as small as 
$10^{-16}$Pa. This technique can be self-calibrated and highly reliable above the  $10^{-14}$Pa
scale.

We thank Juan Hern\'andez-Toro, Jos\'e A. P\'erez-Hern\'andez and Luis Roso for useful discussions. A. P. is supported by the Ram\'on y Cajal programme. 
D. N. acknowledges support from the spanish MINECO through the FCCI. ACI-PROMOCIONA project (ACI2009-1008).


\end{document}